\documentclass[a4paper]{article}
\usepackage{amsmath,pstricks}

\newcommand{\eqa}{\begin{eqnarray}}
\newcommand{\neqa}{\end{eqnarray}}
\newcommand{\equ}{\begin{equation}}
\newcommand{\nequ}{\end{equation}}

\newcommand{\Ref}[1]{(\ref{#1})}

\begin{document}

\title{\bf{LQG propagator: 
III. The new vertex}} 
\author{\normalsize{{Emanuele Alesci${}^{ab}$, Eugenio Bianchi${}^{bc}$, Carlo Rovelli${}^b$}}\\[1mm]
\small{\em ${}^a$ Laboratoire de Physique, ENS Lyon, CNRS UMR 5672, 46 All\'ee d'Italie, 69364
Lyon, EU}\\[-.5mm]
\small{\em ${}^b$Centre de Physique Th\'eorique de Luminy\footnote{Unit\'e mixte de recherche (UMR 6207) du CNRS et des Universit\'es de Provence (Aix-Marseille I), de la Mediterran\'ee (Aix-Marseille II) et du Sud (Toulon-Var); laboratoire affili\'e \`a la FRUMAM (FR 2291).}, Universit\'e de la M\'editerran\'ee, F-13288 Marseille, EU}\\[-.5mm]
\small{\em ${}^c$  Scuola Normale Superiore, Piazza dei Cavalieri 7, I-56126 Pisa, EU}\\[-1mm]}

\date{\small\today}
\maketitle \vspace{-.6cm}

\begin{abstract}

\noindent In the first article of this series, we pointed out a difficulty 
in the attempt to derive the  low-energy behavior of  the graviton two-point function,  
from the loop-quantum-gravity dynamics defined by the 
Barrett-Crane vertex amplitude. Here we show that this difficulty disappears when using the 
\emph{corrected} vertex amplitude recently introduced in the literature.  
In particular, we show that the asymptotic analysis of the new vertex amplitude 
recently performed by Barrett, Fairbairn and others, implies that the 
vertex has \emph{precisely} the asymptotic structure that, in the second article of this series, 
was indicated as the key necessary condition for overcoming the difficulty. 
\end{abstract}

\section{Introduction}

A technique for computing $n$-point functions in a background-independent
context has been introduced in \cite{scattering1, scattering2, scattering3} and developed in 
\cite{Livine:2006it}.  Using this technique, we have found in the first paper of this series 
\cite{I} that the definition of the dynamics of loop quantum gravity defined by the 
Barrett-Crane (BC) spinfoam vertex \cite{Barrett:1997gw} \emph{fails} to give 
the correct graviton propagator in the large-distance limit.  This result has prompted 
a lively search for an appropriate correction of the BC vertex \cite{EPR,L,EPRL,LS,FK}.  
The search has yielded an alternative vertex, given by the square of an $SU(2)$ 
Wigner 15j symbol, contracted with certain natural fusion coefficients 
\cite{EPR,Alesci:2008un}.
The vertex can be defined for general values of the Immirzi parameter $\gamma$ \cite{EPRL}
and can be extended to the Lorentzian case \cite{L, EPRL}. The same 
vertex has also been derived \cite{FK} using the coherent states techniques introduced by 
Livine and Speziale \cite{LS}. For $\gamma<1$ the two techniques yield exactly the same 
theory -- the theory we consider here.

In the second article of this series \cite{II}, we argued that the correct graviton propagator in 
the large-distance limit \emph{could} be obtained only \emph{if} the vertex had a certain asymptotic form. The asymptotic analysis of the new vertex amplitude has been recently performed by
John Barrett, Winston Fairbairn, and their collaborators \cite{Barrett}.  Here we show that the result of the Barrett-Fairbairn analysis implies that the new vertex has precisely the asymptotic form guessed in the second article of this series, and therefore it resolves the difficulty of the old Barrett-Crane vertex. 

This paper is not self-contained.  It is based on the two previous papers of this series  \cite{I,II}, where all relevant definitions are given.   For an introduction to the formalism,
see \cite{scattering3}; for a general introduction to background independent
loop quantum gravity \cite{lqg2}, see \cite{lqg1,lqg3}.

\section{Conditions on the vertex asymptotics}

The quantity on which we focus is the (non-gauge-invariant) euclidean propagator $G^{\mu\nu\rho\sigma}(x,y)=\langle 0|h^{\mu\nu}(x)h^{\rho\sigma}(y)|0\rangle$, where 
 $|0\rangle$ is a vacuum state peaked on the flat euclidean metric  $\delta^{\mu\nu}$, and 
$h^{\mu\nu}(x)$ is the difference between the gravitational quantum field and its euclidean background\footnote{More precisely, the proper quantity to consider is $
\langle 0|g^{\mu\nu}(x)g^{\rho\sigma}(y)|0\rangle-
\langle 0|g^{\mu\nu}(x)|0\rangle \langle 0|g^{\rho\sigma}(y)|0\rangle$, which shows that the
information about the background is \emph{entirely} contained in the state; but in this letter we follow the standard practice for simplicity.} value $\delta^{\mu\nu}$. Let $L$ be the distance between $x$ and $y$ (in the flat euclidean metric). 
 Choose a regular 4-simplex with two boundary tetrahedra $n$ and $m$ centered at the points $x$ and $y$; the indices $i,j,k,l,m,n,...=1,...,5$ label the five tetrahedra bounding the 4-simplex.  Define ${\mathbf G} _{n,m}^{\scriptscriptstyle ij,kl}(L)=
G^{\mu\nu\rho\sigma}(x,y)(n^{\scriptscriptstyle(i)}_n)_\mu (n^{\scriptscriptstyle(j)}_n)_\nu (n^{\scriptscriptstyle(k)}_m)_\rho (n^{\scriptscriptstyle(l)}_m)_\sigma$, where  $n_m^{\scriptscriptstyle(k)}$ is the normal one-form to the triangle bounding the tetrahedra $m$ and $k$,  in the hyperplane defined by the tetrahedron $m$ (with $|n|$ equal to the area of the triangle).  Clearly, knowing  ${\mathbf G} _{n,m}^{\scriptscriptstyle ij,kl}(L)$ is the same as knowing $G^{\mu\nu\rho\sigma}(x,y)$.   Following \cite{scattering1,scattering2,scattering3}, ${\mathbf G} _{n,m}^{\scriptscriptstyle ij,kl}(L)$  can be computed 
in a background  independent context as the scalar product 
\begin{equation}
{\mathbf G} _{{\mathbf q}\, n,m}^{\scriptscriptstyle ij,kl} = \langle W | \big(E^{\scriptscriptstyle(i)}_n \cdot E^{\scriptscriptstyle(j)}_n-n_n^{\scriptscriptstyle(i)}\cdot n_n^{\scriptscriptstyle(j)}\big)
\big(E^{\scriptscriptstyle(k)}_m  \cdot E^{\scriptscriptstyle(l)}_m-n_m^{\scriptscriptstyle(k)}\cdot n^{\scriptscriptstyle(l)}_m\big) |\Psi_{\mathbf q} \rangle.
\label{partenza0}
\end{equation}
Here $ \langle W |$ is the boundary functional, which can be intuitively understood as the 
path integral of the Einstein-Hilbert action on a finite spacetime region $\cal R$, with given
boundary configuration.   The operator $E_n^{\scriptscriptstyle(i)}$ is the triad operator at the point $n$, contracted with $n_n^{\scriptscriptstyle(i)}$.  $|\Psi_{\mathbf q} \rangle$
is a state on the boundary of $\cal R$, peaked on a given classical boundary (intrinsic and 
extrinsic)  geometry $\mathbf q$.
Fixing such a boundary geometry is equivalent to fixing a background metric $g$ in the 
interior, where $g$ is the solution of the Einstein equations with boundary data $\mathbf q$. 
The existence of such a background metric is part of the definition of the propagator:
the propagator is indeed a measure of the fluctuations around a given background. The (intrinsic and extrinsic) boundary geometry chosen in  \cite{scattering1, scattering2, scattering3,Livine:2006it,I} is that of the boundary of a \emph{regular} four-simplex, immersed in $R^4$. 
The classical Ricci flat bulk metric $g$ determined by these boundary data is obviously the flat metric, thus allowing a comparison with the free-graviton propagator of the theory linearized
around flat space.
It is convenient to write
\begin{equation}
|\Psi_{{\mathbf q}\, n,m}^{\scriptscriptstyle ij,kl}\rangle = \big(E^{\scriptscriptstyle(i)}_n \cdot E^{\scriptscriptstyle(j)}_n-n_n^{\scriptscriptstyle(i)}\cdot n_n^{\scriptscriptstyle(j)}\big)
\big(E^{\scriptscriptstyle(k)}_m  \cdot E^{\scriptscriptstyle(l)}_m-n_m^{\scriptscriptstyle(k)}\cdot n^{\scriptscriptstyle(l)}_m\big) |\Psi_{\mathbf q} \rangle
\label{stato}
\end{equation}
so that
\begin{equation}
{\mathbf G} _{{\mathbf q}\, n,m}^{\scriptscriptstyle ij,kl} = \langle W 
|\Psi_{{\mathbf q}\, n,m}^{\scriptscriptstyle ij,kl}\rangle.
\label{partenza1}
\end{equation}

We are interested in the value of \Ref{partenza1} on a triangulation formed by a single 4-simplex,
or,  equivalently,  to first order in the group-field-theory \cite{GFT} expansion parameter, and in the limit in which the boundary surface (whose size is 
determined by  $\mathbf q$) is large. On the physical interpretation of this approximation, see 
\cite{EPR}.  To first order, the leading contribution to 
$W$ has support only on spin networks with a 4-simplex graph. If 
${\mathbf j}=(j_{nm})$ and ${\mathbf k}=(k_n)$
are, respectively, the ten spins and the five (spins of the virtual links labeling the 
five) intertwiners that color this graph, then in this
approximation (\ref{partenza1}) reads
\begin{equation}
{\mathbf G} _{{\mathbf q}\, n,m}^{\scriptscriptstyle ij,kl} 
 = \sum_{{\mathbf j}, {\mathbf k}}\ W({\mathbf j}, {\mathbf k})
  \ \Psi_{{\mathbf q}\, n,m}^{\scriptscriptstyle ij,kl}({\mathbf j}, {\mathbf k})
  \label{partenza} 
\end{equation}
To this order, $W$ is just determined by the amplitude of a single vertex (up to some
normalization factors that are irrelevant here). Since $\Psi_{{\mathbf q}\, n,m}^{\scriptscriptstyle ij,kl}({\mathbf j}, {\mathbf k})$ is peaked on large values $j_{nm}=j_0$ and $k_n=k_0$ of ${\mathbf j}$ and  ${\mathbf k}$, the propagator depends only on the asymptotic (large ${\mathbf j}$ and large ${\mathbf k}$) behavior of the vertex, or, more precisely, on the behavior of $ W(j_0+\delta{\mathbf j}, k_0+\delta{\mathbf k})$ for large ($j_0$, $k_0$) and small  $(\delta j_{nm}, \delta k_n)=(j_{nm}-j_0,  k_n-k_0)$. 
 
In the second paper of this series \cite{II}, we showed that \emph{if} the vertex $W$ has a certain 
asymptotic structure, described below, \emph{then} the boundary state can be appropriately chosen
to give the correct propagator in the large distance limit.  Expanding $W$ to second order in the fluctuations, we write 
\begin{equation} 
	W(j_0+\delta\mathbf{j},k_0+\delta \mathbf{k}) \sim N 
	e^{{i}G(\delta{j_{nm}},\delta{k_n})} 
	e^{i\phi_{nm}\delta{j}_{nm}+i\phi_{n}\delta{k}_{n}} \ \ +c.c.
	\label{caso11}
\end{equation}
Here $N$ is a slowly varying function,  $G$ is a quadratic form in its 15 arguments, which scales as $1/j_0$.   The key quantity for us is the 15d vector $(\phi_{nm},\phi_n)$ that determines the first order variation of $W$ around $(j_0,k_0)$: it is the frequency of the rapidly oscillating phase factor around this point.  The result of \cite{II} is that if $W$ has the form (\ref{caso11}), with ``appropriate'' values of  $(\phi_{nm},\phi_n)$ of the phases, then we obtain the correct graviton propagator.     ``Appropriate'' means here that these phases must cancel corresponding phases in the boundary state. 

To explain this point in detail, let us pause one moment and consider the general situation in quantum mechanics.  In general, in quantum mechanics, the frequency of this rapidly oscillating factor codes the classical equations of motion, and therefore gives the semiclassical limit of the dynamics.  For instance, the propagator of a free particle in a time $t$ between two points $x_0$ and $y_0$ behaves like 
\begin{equation} 
W_t(x_0+\delta x, y_0+\delta y)\equiv \langle x_0+\delta x |e^{i Ht}|  y_0+\delta y \rangle \sim e^{i(p_x\delta x-p_y\delta y)}
\end{equation}
where the frequency of the oscillation is precisely the momentum of the classical trajectory going from $x_0$ to $y_0$, that is 
\begin{equation}
p_x=p_y=m\ \frac{y_0-x_0}{t}.
\label{free}
\end{equation}
 To see how the semiclassical limit works, consider sandwiching the propagator between two wave packets centered in $x_0$ and $y_0$ respectively, and with oscillating phases having values $P_x$ and $P_y$ respectively. Then the amplitude is suppressed by the rapidly oscillating phase factor unless  $P_x=m(y_0-x_0)/t$ and $P_y=m(y_0-x_0)/t$. That is: a semiclassical wave packet propagates from $x_0$ to $y_0$ in a time $t$ only if it starts (and ends) with the correct momentum (velocity). If we propagate further for an addition time, destructive interference is avoided only along the classical trajectory. 

More precisely, the classical equations of motion are normally interpreted as giving $y_0$ and $p_y$ given $x_0$ and  $p_x$; but they can equally be interpreted as giving the momenta $p_x$ and $p_y$ given $x_0$ and  $y_0$.  This second interpretation is the one that generalizes naturally in a general covariant context \cite{lqg1}.  Under this second interpretation, the classical equations of motion determine the momenta, (and hence the gluing conditions for paths) given initial and final positions (or, more in general, given boundary configuration variables).  The semiclassical limit of quantum mechanics determines these momenta as the phases of the propagator.  Thus, the value of these phases codes the classical limit of the theory.  Another way of seeing the same point is to observe that in the semiclassical approximation the propagator is given by the exponential of the classical action along a given classical solution of the equations of motion; that is, by the Hamilton function \cite{lqg1}. When varying a boundary configuration variable, the variation of the Hamilton function gives the momenta. 

In the expression \Ref{caso11}, the 15 factors $\phi=(\phi_{nm},\phi_n)$ determine the semiclassical behavior of the theory near the configuration $(j_0,k_0)$.  These are determined by the geometry on which the boundary state is peaked (like the momenta $p_x, p_y$ are determined by $x_0$ and $y_0$ in \Ref{free}).  In the case we are considering, the boundary state is chosen to be peaked on the geometry of a regular four-simplex in a flat spacetime. 

The phase factor of the  ${j}_{nm}$ variables in the boundary state gives the mean value of the variable canonically conjugate to the variables ${j}_{nm}$. The variables ${j}_{nm}$ represent the areas of the ten faces in the four-simplex. The variables canonically conjugate to these were identified in \cite{scattering2} as the momentum variables conjugate to the triangle areas, namely the 4d dihedral angle between adjacent tetrahedra, as is the case in Regge calculus. For a regular 4-simplex, this gives 
\begin{equation}
\cos\phi_{nm} = -\frac14.
\label{1}
\end{equation}
Remarkably, the BC vertex has precisely this phase factor.

The situation with the angles $\phi_n$ is more delicate, and crucial. The intertwiner $k_n$ is the quantum number of a dihedral angle $\beta$ between two chosen faces of the tetrahedron $n$.  The semiclassical boundary state will peak it on some value $k_0$. But the semiclassical state must peak also the value of the quantum number of the dihedral angle between {\em any two other} faces of the same tetrahedron $n$, and these are quantities that do not commute with $\beta$  \cite{tetraedro}. The only possibility for having all angles peaked is to have a properly chosen semiclassical state. The phase dependence of the state on $k_n$ determines where the other angles are peaked (like the phase dependence of a wave packet depends where the packet is peaked in momentum space).  In \cite{tetraedro2} is was shown that in order to have all angles peaked on the proper values for a regular tetrahedron, the value of this phase must be
\begin{equation}
 \phi_n = -\frac\pi2. 
\label{2}
\end{equation}
(For a general boundary geometry, the values \Ref{1} generalize to appropriate functions of this geometry \cite{Eugenio}.)  The BC vertex does \emph{not} have this phase dependence, and this was the key detail that made the derivation of the propagator impossible in \cite{I}. In \cite{II} we observed that this phase dependence is needed for having a boundary state yielding the correct propagator.  A hypothetical vertex characterized by an asymptotic behavior around large $j_0, k_0$ being like the BC vertex (namely asymptotic to the Regge action) but also having such a phase dependence was therefore \emph{guessed} in  \cite{II}, and shown to yield the correct propagator with an appropriate boundary state.

Here we show that the new vertex has precisely this asymptotic structure, with the phase \Ref{2}.  This is a consequence of the asymptotic analysis of the vertex, recently completed by Barrett, Fairbairn {\em et.\,al.}, and which we summarize below.  

\section{The asymptotic of the new vertex}

We state here the result of the asymptotic form of the vertex found by Barrett, Fairbairn  {\em et.\,al.}.  This is given in a different basis for the intertwiners: the overcomplete basis $|n \rangle$ of the coherent state, introduced by Livine and Speziale \cite{LS}.  Here $n$ is a quadruplet of unit-length vectors in $R^3$ (whose geometrical interpretation is the normal of the faces of a tetrahedron).  The matrix elements  $\langle k | n \rangle$  of this change of basis are defined in   \cite{LS} and their asymptotic behavior has been studied in detail by one of us in \cite{Eugenio}. 

The result of the asymptotic analysis of Barrett, Fairbairn {\em et.\,al.} is the following \cite{Barrett}. In the large $\bf j$ limit, the vertex defined in \cite{EPRL} behaves as follows:
\begin{equation}
	W(\mathbf{j},{\bf n}):=\langle W|{\bf j},{\bf n}\rangle\approx N e^{-\imath S(\mathbf{j},{\bf n})} \mu(\mathbf{j},{\bf n}) +c.c.
	\label{John}
\end{equation}
Here ${\bf n}=(n_n)$ is the family of the five quadruplets of $R^3$ vectors forming the basis of the five intertwiner spaces.  $N$ is a slowly varying factor.  $\mu(\mathbf{j},{\bf n})$ is a factor that is strongly suppressed when  $(\mathbf{j},{\bf n})$ are \emph{not} ``consistent''. Here ``consistent'' means that there exist a 4-simplex imbedded in $R^4$ whose triangles have the areas determined by the spins  $\mathbf{j}$ and whose tetrahedra have the (3d) geometry determined by the normals $\bf n$.  When   $(\mathbf{j},{\bf n})$ are consistent, $S(\mathbf{j},{\bf n})$ is the Regge action \cite{Regge:1961px} of such a geometrical 4-simplex (more precisely, the Dittrich-Speziale action  \cite{Dittrich:2008va}), divided by $8\pi \hbar G_{\rm Newton}$. This is obtained identifying $j_{nm}$ with ${1}/{8\pi \gamma \hbar G_{\rm Newton}}$ ($\gamma$ is the Immirzi parameter) times the area of the boundary triangle -- this being the proper identification in the theory, for large $j_{nm}$ \cite{EPRL}.  This very remarkable result is obtained via a saddle point evaluation of the vertex amplitude defined in \cite{EPRL}, written as an integral over copies of $SU(2)$, with techniques derived from  \cite{Barrett:1998gs}; see also \cite{CF}. 

What we now need to show is that this result implies that $W$ has the correct asymptotic form (\ref{caso11}, \ref{2}).  We can discard the $c.c.$ term in both expressions \Ref{caso11} and \Ref{John}, since we know from \cite{scattering2} that the boundary state selects only one of the two $c.c.$ terms.  What we need to do is simply to transform \Ref{John} to the intertwiner basis $\bf k$, that is, compute
\begin{equation}
W({\mathbf j},{\mathbf k})=\int d{\bf n} \ 
\langle W|{\bf j},{\bf n}  \rangle\langle {\bf n},  |{\mathbf k}\rangle.
\end{equation}
We are only interested in this expression in the vicinity of $(j_0,k_0)$. More precisely, we are 
specifically interested in the first order variation of $W$ when varying $k$:
\begin{equation}
W(j_0,k_0+\delta \mathbf{k})
=\int d{\bf n} \ 
\langle W|j_0, {\bf n} \rangle\langle {\bf n},  |k_0+\delta \mathbf{k}  \rangle.
\end{equation}
Since $j_0$ is large, we can use the Barrett-Fairbairn result \Ref{John}. Inserting in the last equation gives
\begin{equation}
W(j_0,k_0+\delta \mathbf{k})
=\int d{\bf n} \ 
N e^{-\imath S(\mathbf{j_0},{\bf n})} \mu(\mathbf{j_0},{\bf n}) 
\langle {\bf n},  |k_0+\delta \mathbf{k}  \rangle.
\label{aas}
\end{equation}
Now, the $\mu(\mathbf{j_0},{\bf n}) $ factor peaks the integral on those $\bf n$ that define a 4-simplex with all equal areas $j_0$.  But the geometry of a four simplex is entirely determined by these ten areas (up to possible discrete degeneracy that we disregard here).  Hence the $\bf n$ that contribute to the integral are only those characterizing regular tetrahedra.  
These still have a $SU(2)^5$ multiplicity, because being normal to the faces of a \emph{regular} tetrahedron defines the quadruplet of vectors ${\bf n}_0$ only up to a global $SO(3)$ rotation, but this rotation does not affect \Ref{aas} since $S$ and the intertwiner states $| k \rangle$ are $SU(2)$ invariant.   Hence, trivially integrating out this subgroup we can write 
\begin{equation}
W(j_0,k_0+\delta \mathbf{k})
= 
N e^{-\imath S(\mathbf{j_0})} \langle {\bf n}_0,  |k_0+\delta \mathbf{k}  \rangle.
\end{equation}
where ${\bf n}_0$ is an arbitrary set of five quadruplets of vectors, each normal to the faces of a {\em regular} four-simplex.  

We now need the asymptotic expression for the matrix elements of the change of basis.  It is shown in \cite{Eugenio} that when $\bf n$ are the vectors ${\bf n}_0$ normal to the faces of a \emph{regular} tetrahedron, the state $|\bf n \rangle$ can be shown to converge to the coherent tetrahedron state defined in \cite{tetraedro2}. This behaves like 
\begin{equation}
\langle k | {\bf n}_0\rangle \sim e^{i \frac{\pi}2 k}
\label{as}
\end{equation}
for large spins.   Inserting this we have 
\begin{equation}
W(j_0,k_0+\delta \mathbf{k}) \sim N e^{-\imath S(\mathbf{j_0})}\  e^{i \frac{\pi}2 k}.
\end{equation}
The important result here is the appearance of the correct $\pi/2$ factor in the phase, which was missing in the BC vertex.  This is precisely the phase \Ref{2} we were looking for.   Thus, we new vertex has the asymptotic behavior that was \emph{guessed} in \cite{II}, in order to yield the graviton propagator. This is our main observation. 

It is also instructive to analyze the first order dependence on the spins. Keeping now the intertwiner fixed, we have 
\begin{eqnarray}
W(j_0+\delta \mathbf{j},k_0)
&=&\int d{\bf n} \ 
\langle W|j_0+\delta \mathbf{j}, {\bf n} \rangle\langle {\bf n},  |k_0\rangle.
\nonumber \\ 
&=&\int d{\bf n} \ 
N e^{-\imath S(\mathbf{j_0+\delta \mathbf{j}},{\bf n})} \mu({j_0}+\delta \mathbf{j},{\bf n}) 
\langle {\bf n},  |k_0\rangle.
\end{eqnarray}
Taking the factor $\mu$ of the saddle point approximation to behave like a delta function in the large $j$ limit in which we are, we obtain 
\begin{eqnarray}
W(j_0+\delta \mathbf{j},k_0)
&=&
N e^{-\imath S(\mathbf{j_0+\delta \mathbf{j}},{\bf n(j_0+\delta \mathbf{j})}} 
\langle {\bf n}(j_0+\delta \mathbf{j})  |k_0\rangle
\nonumber \\ 
&=&
N e^{-\imath S_{Regge}(\mathbf{j_0+\delta \mathbf{j}})}\  e^{i {\bf \Phi({\bf n}(j_0+\delta \mathbf{j})} )k_0},
\end{eqnarray}
where $S_{Regge}(\mathbf{j)}\equiv S(\mathbf{j,\mathbf{n(j))}}$ and $\mathbf{n(j)}$ are the 
$\mathbf{n}$ consistent with $\mathbf{j}$ in the neighborhood of $\mathbf{n}_0$. The angle
$\bf\Phi$ is determined in \cite{Eugenio} as the angle between the two paired edges of the tetrahedron.  Expanding this gives 
${\bf \Phi}({\bf n}(j_0+\delta \mathbf{j}))={\bf \Phi}({\bf n}(j_0))+{\bf\Phi'}\delta \mathbf{j}=\frac\pi2+ {\bf\Phi'}\delta \mathbf{j}$.  The first order variation of the  $S_{Regge}$ action around a flat 4 simplex gives precisely the $\phi_{nm}$ phases satisfying \Ref{1}. But these get corrected by the  additional term ${\bf \Phi'} k_0$ coming from the matrix elements, which must be taken into account in fixing the boundary state.   If we expand to the next order, the full quadratic part of \Ref{caso11} is determined by the Hessian of the Regge action.

\section{Conclusions}

What we have shown here is only that the obstacle that prevented the Barrett-Crane vertex to yield the proper graviton propagator is resolved by the new vertex.  Clearly it is now necessary to restart from scratch the calculation of the graviton propagator using the new vertex, and check that everything works properly.  For this, it seems clear that the best basis to use in the intertwiner spaces is the Livine-Speziale coherent state basis. The disadvantage of using a badly over-complete basis is overcome by the advantage of having a basis that respects the symmetries of the four-simplex, thus avoiding the complications due to the need to choose a pairing in picking a virtual-spin basis.  The full calculation of the graviton two-point function using the Livine-Speziale basis is in course, and will be presented elsewhere.  Calculations of higher-$n$,  $n$-point functions and higher-order terms of the propagator are also in course. 

A number of issues need better clarification before we can say that we understand the low energy limit of loop quantum gravity. Among these the role of gauge invariance \cite{ClaudioElena} and  finiteness \cite{finiteness}.   Nevertheless, we see good reasons for optimism.  The new vertex has been introduced in \cite{EPR, EPRL} only as an attempt to give the intertwiners a dynamics. Whether this dynamics was correct at low-energy remained unclear during the last year, lacking the asymptotic analysis of the vertex. Remarkably, this analysis turns out to give precisely the intertwiner dependence that was previously indicated as the one hoped for.

%%%%%%%%%%%%%%%%%%%%%%%%%%%%%%%%%%%%%%%%%%%

\end{document}